\documentclass[fleqn,usenatbib]{mnras}
\usepackage{graphicx}	
\usepackage{amsmath}	
\usepackage[dvipsnames]{xcolor}  
\usepackage{amssymb}	
\usepackage{newtxtext,newtxmath}    
\usepackage{siunitx}
\usepackage{ulem}
\usepackage{macros_vdb} 

\usepackage{macros_vdb}
\usepackage{siunitx}

\usepackage{graphicx}	
\usepackage{amsmath}	
\usepackage{amssymb}	

\defcitealias{Hayashi.etal.03}{H03}
\defcitealias{Penarrubia.etal.10}{P10}

\title[Subhalo density profile evolution]
      {The tidal evolution of dark matter substructure -- I. Subhalo density profiles}

\author[S. B. Green and F. C. van den Bosch]{
        Sheridan~B.~Green$^{1}$\thanks{E-mail: \href{mailto:sheridan.green@yale.edu}{sheridan.green@yale.edu} (SBG)}\thanks{NSF Graduate Research Fellow} and Frank~C.~van den Bosch$^{1,2}$
\vspace*{8pt}
\\
$^{1}$Department of Physics, Yale University, P.O. Box 208120, New Haven, CT 06520-8120\\
$^{2}$Department of Astronomy, Yale University, P.O. Box 208101, New Haven, CT 06520-8101\\
}

\date{}

\pubyear{2019}

\begin{document}
\label{firstpage}
\pagerange{\pageref{firstpage}--\pageref{lastpage}}

\maketitle


\begin{abstract}
   Accurately predicting the abundance and structural evolution of dark matter subhaloes is crucial for understanding galaxy formation, modeling galaxy clustering, and constraining the nature of dark matter. Due to the nonlinear nature of subhalo evolution, cosmological $N$-body simulations remain its primary method of investigation. However, it has recently been demonstrated that such simulations are still heavily impacted by artificial disruption, diminishing the information content on small scales and reducing the reliability of all simulation-calibrated semi-analytical models. In this paper, we utilize the recently released DASH library of high-resolution, idealized simulations of the tidal evolution of subhaloes, which are unhindered by numerical overmerging due to discreteness noise or force softening, to calibrate an improved, more-accurate model of the evolution of the density profiles of subhaloes that undergo tidal heating and stripping within their host halo. By testing previous findings that the structural evolution of a tidally truncated subhalo depends solely on the fraction of mass stripped, independent of the details of the stripping, we identify an additional dependence on the initial subhalo concentration. We provide significantly improved fitting functions for the subhalo density profiles and structural parameters ($V_\mathrm{max}$ and $r_\mathrm{max}$) that are unimpeded by numerical systematics and applicable to a wide range of parameter space. This model will be an integral component of a future semi-analytical treatment of substructure evolution, which can be used to predict key quantities, such as the evolved subhalo mass function and annihilation boost factors, and validate such calculations performed with cosmological simulations.
\end{abstract}




\section{Introduction}

In the $\Lambda$ cold dark matter ($\Lambda$CDM) cosmological model of structure formation, primordial density perturbations with a scale-invariant power spectrum collapse to form virialized haloes. Due to the negligible free-streaming velocities of CDM, haloes \rev{form} on all scales, with smaller perturbations collapsing earlier and subsequently assembling from the bottom up to form more massive haloes. Since 1997, cosmological $N$-body simulations have shown that the dense, inner regions of these smaller haloes continue to live on as subhaloes within their hosts after having been accreted \citep{Tormen.etal.97, Moore.etal.98b, Ghigna.etal.98}, and these subhaloes themselves host sub-subhaloes, and so on, forming a complete hierarchy of substructure \citep{Gao.etal.04, Springel.etal.08, Giocoli.etal.10}. As these subhaloes orbit their hosts, they are subjected to various forces that work to disrupt them, including dynamical friction, tidal stripping and impulsive heating due to the host, and harassment by other substructure \citep[e.g.,][]{MBW10, vdBosch.etal.2018a}.

The statistics of dark matter (DM) substructure are sensitive to the underlying DM model. In particular, the DM thermal velocity sets the cutoff scale for low-mass haloes, which in turn impacts the abundance of substructure \citep[e.g.,][]{Knebe.etal.08, Lovell.etal.14, Colin.etal.15, Bose.etal.17}, and the (potentially nonzero) cross-section for DM self-interaction can core out the otherwise cuspy slopes of subhalo inner density profiles, making them less resilient to the strong tidal forces of the host halo \citep[e.g.,][]{Burkert2000, Vogelsberger.etal.12, Rocha.etal.13}. The primary observational techniques used to probe the properties of DM substructure include gravitational lensing \citep[e.g.,][]{Dalal.Kochanek.02, Keeton.Moustakas.09, Vegetti.etal.14, Hezaveh2016, Gilman.etal.19}, gaps in stellar streams \citep[e.g.,][]{Carlberg2012, Ngan2014, Erkal.etal.16}, and indirect detection via DM annihilation and decay signals \citep[e.g.,][]{Strigari.etal.07, Pieri.etal.08, Hayashi2016, Hiroshima2018, Delos2019}. Furthermore, since \rev{satellite galaxies are expected to reside within some fraction of the DM subhaloes}, the demographics of DM substructure has a direct correspondence to that of satellite galaxies \citep[e.g.,][]{Vale.Ostriker.06, Hearin.etal.13, Behroozi.etal.13c, Newton.etal.18}, which ultimately impacts small-scale clustering statistics \citep[e.g.,][]{Benson.etal.01, Berlind.etal.03, Kravtsov.etal.04, Campbell.etal.18}. Thus, being able to accurately predict the abundance and structural evolution of DM subhaloes is paramount for using astrophysics to study the particle nature of dark matter. 

Due to its high nonlinearity, a purely analytical description of subhalo evolution is impossible, even in the most idealized of circumstances \citep[for a detailed discussion, see][]{vdBosch.etal.2018a}. Hence, the primary method employed for studying the demographics of DM substructure has been, and remains, cosmological $N$-body simulations. \rev{Prior to the late 1990s, numerical simulations did not yet have sufficient mass and force resolution to resolve surviving populations of subhaloes \citep{Moore.etal.96a, Klypin.etal.99a}. As increased computational power has enabled access to ever higher resolutions, many} convergence tests have \rev{since} been performed to validate the results of \rev{more recent} $N$-body simulations, demonstrating consistent subhalo mass functions above a resolution limit of 50-100 particles \citep[e.g.,][]{Springel.etal.08, Onions.etal.12, Knebe.etal.13, vdBosch.Jiang.16, Griffen.etal.16}; however, mass function convergence is only a necessary, but not sufficient, condition to guarantee the physical correctness of numerical simulations. \Citet{vdBosch.17} showed that the complete disruption of subhaloes occurs very frequently in state-of-the-art simulations, with a mass function of disrupted subhaloes that is identical to that of the surviving population. The inferred disruption rate implies that roughly 65\% of subhaloes accreted around $z=1$ are disrupted by $z=0$ \citep{Han.etal.16, Jiang.vdBosch.17}. Some authors have argued that complete disruption is a physical consequence of tidal heating and/or tidal stripping \citep{Hayashi.etal.03, Taylor.Babul.04, Klypin.etal.15}. However, \citet{vdBosch.etal.2018a} demonstrated that neither tidal heating nor tidal stripping are independently sufficient to completely disrupt CDM subhaloes, a result consistent with the idealized, high-resolution numerical simulations of \citet{Penarrubia.etal.10}. \Citet{vdBosch.etal.2018b} ran a suite of similar, idealized numerical experiments, finding that subhalo disruption in $N$-body simulations is largely due to two key numerical details: (i) discreteness noise due to insufficient particle resolution and (ii) inadequate force softening. \rev{The optimal force softening criteria put forth by \citet{vdBosch.etal.2018b} have since been corroborated by \citet{Ludlow2019} and are in good agreement with the criteria of \citet{Zhang2019}.}

This artificial subhalo disruption may have substantial consequences across cosmology and astrophysics. For example, in small-scale clustering analysis, the uncertainty due to disruption reduces the predictive power of methods such as subhalo abundance matching \citep[e.g.,][]{Vale.Ostriker.06, Conroy.etal.06, Guo.etal.10, Hearin.etal.13}, while the reduced abundance of substructure implies that dark matter annihilation boost factors \citep[e.g.,][]{Bergstrom.etal.99, Ando.etal.19} may be substantially underestimated. The all-important, outstanding question is to what extent this artificial disruption impacts the subhalo mass and/or velocity function predicted by cosmological simulations. The work of \citet{vdBosch.etal.2018b} suggests that the answer is unlikely to come from numerical simulations, as there is no obvious way to circumvent the numerical issues. Instead, we may hope to gain some insight from semi-analytical models of the build-up and evolution of dark matter substructure \citep[e.g.,][]{Taylor.Babul.01, Penarrubia.Benson.05, Zentner.etal.05, vdBosch.etal.05, Kampakoglou.Benson.07, Gan2010, Pullen.etal.14}. The problem, though, is that the lack of a complete theory of tidal evolution implies that these semi-analytical models need to be calibrated, which is typically done by tuning the model to reproduce the subhalo mass functions inferred from cosmological $N$-body simulations. This obviously implies that the models inherit the shortcomings of the simulations. The main goal of this paper is to present a model of the evolution of subhalo density profiles that circumvents this catch-22 situation. 

\begin{figure*}
    \centering
    \includegraphics[width=\linewidth]{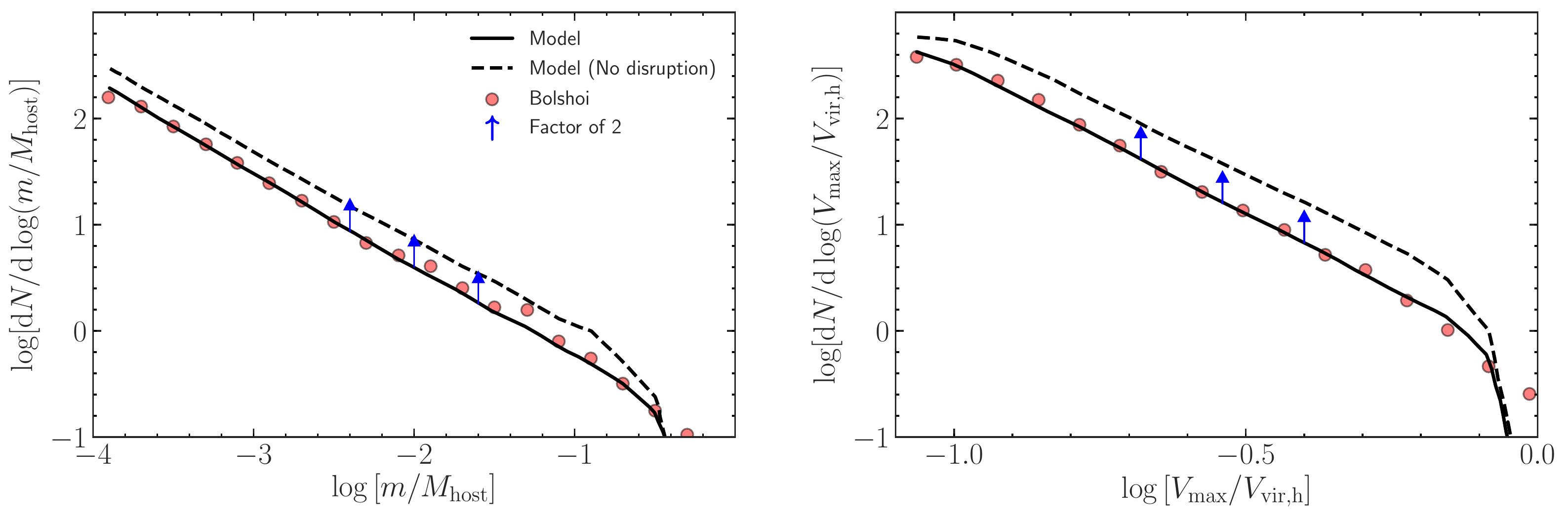}
    \caption{Subhalo mass ({\it left}) and velocity ({\it right}) functions for host haloes with masses in the range $14.0 \leq \log[M_\rmh/\Msunh] \leq 14.5$. Symbols indicate the results obtained from the {\tt Bolshoi} simulation \citep[][]{Klypin.etal.11}, whereas the solid lines are the results obtained from the semi-analytical model of \citet{Jiang.vdBosch.16}. The latter includes models for subhalo mass loss and subhalo disruption that have been tuned to specifically reproduce the subhalo mass and velocity functions of the {\tt Bolshoi} simulation. The dashed lines show the predictions of the same model, but with subhalo disruption turned off. The inference is that if the majority of subhalo disruption is artificial, as claimed by several recent studies \citep[][]{Penarrubia.etal.10, vdBosch.etal.2018a, vdBosch.etal.2018b}, state-of-the-art cosmological simulations may under-predict the abundance of subhaloes by as much as a factor of two (blue arrows). See text for a more detailed discussion.}
    \label{fig:shmf}
\end{figure*}

Before describing our methodology, though, it is insightful to try to estimate how big of an impact artificial disruption may potentially have. We can do so using the semi-analytical model of \citet{Jiang.vdBosch.16}, which combines halo merger trees with simple models of the tidal evolution of subhaloes, to predict the evolved subhalo mass and velocity functions of dark matter substructure \citep[see][]{Jiang.vdBosch.17}. The model treats both mass stripping as well as subhalo disruption, the efficiencies of which are calibrated to reproduce the results of the high-resolution {\tt Bolshoi} simulation \citep{Klypin.etal.11}. The left- and right-hand panels of Fig.~\ref{fig:shmf} plot the subhalo mass and velocity functions, respectively. The solid circles indicate the results from the {\tt Bolshoi} simulation for present-day host haloes with masses in the range $14.0 \leq \log[M_\rmh/\Msunh] \leq 14.5$, while the solid line is the model prediction from \citet{Jiang.vdBosch.16}. Since the latter is calibrated against the former, it should not come as a surprise that the model fits the simulation data well. However, as discussed at length in \citet{Jiang.vdBosch.16}, crucial for this success is the separate treatment of subhalo disruption. We can now use this model to predict what the subhalo mass and velocity functions would look like under the assumption that {\it all} disruption is artificial. To that extent, we rerun the same model, this time turning off disruption; in this case, subhaloes continue to experience mass loss rather than fully disrupt. The resulting mass and velocity functions are indicated by the dashed curves. Clearly, artificial disruption does not merely impact the mass/velocity functions at the low mass end, close to the resolution limit of the simulation; rather, the mass and velocity functions are boosted globally by factors of $\sim 2$ and $\sim 2.5$, respectively. If these admittedly crude predictions are even remotely correct, the implications are far-reaching. It suggests that state-of-the-art cosmological simulations systematically under-predict the abundance of substructure by as much as a factor of two, which, interestingly, is precisely what is needed to solve the `galaxy clustering crisis' in subhalo abundance matching \citep{Campbell.etal.18}. At the very least, these results signal the need to carefully examine the tidal evolution of subhaloes in more detail, which is the core-motivation behind the study presented here.

Semi-analytical models of the build-up and evolution of dark matter substructure consist of three main ingredients: (i) a halo merger tree, which quantifies the subhalo masses and redshifts at accretion, (ii) a model of the orbital evolution, including dynamical friction \rev{and self-friction (Miller et al., in prep.)}, and (iii) a model that describes how the mass and density profile of a subhalo evolves subject to the tidal forces that it experiences. \rev{Semi-analytical merger tree algorithms are calibrated using merger histories from cosmological simulations, which depend on the halo properties at infall and are therefore less sensitive to the effects of artificial disruption than the evolution of individual subhaloes. On the other hand, the evolution of the subhalo density profile} typically requires a model of how the bound mass of the subhalo evolves with time and how this affects the subhalo's density profile. Neither of these can be treated analytically from first principles, and the models therefore typically rely on parametrized treatments that somehow need to be calibrated. In order to prevent the catch-22 situation eluded to above, in \citet{Ogiya2019} we introduced the Dynamical Aspects of SubHaloes (DASH) database, a large library of idealized, high-resolution $N$-body simulations of the tidal evolution of individual subhaloes. These simulations cover a wide range of relevant subhalo parameters (i.e., orbital energy and angular momentum at infall and halo concentrations) and are evolved with sufficient numerical resolution to assuage the impact of discreteness noise and insufficient force softening.  As a next step towards building a more accurate semi-analytical treatment of dark matter substructure evolution, the present paper sets out to develop a new model of the tidal evolution of the subhalo density profile, calibrated against DASH and therefore unimpeded by numerical artifacts, that is applicable to a far wider range of subhalo parameter space than that of previous works \citep[][]{Hayashi.etal.03, Penarrubia.etal.10, Drakos2017}.

This paper is organized as follows: \S\ref{sec:dash} provides an overview of the DASH simulation database. In \S\ref{sec:esdp}, we describe the methods used for building and calibrating our model of the evolved subhalo density profile and then quantify the model's capability of reproducing simulated subhalo density profiles. In \S\ref{sec:struct_evol}, we demonstrate the model's performance at capturing the evolution of the subhalo structural parameters, $V_\mathrm{max}$ and $r_\mathrm{max}$. Lastly, in \S\ref{sec:discussion}, we summarize the results and discuss future work.

\section{The DASH Database}\label{sec:dash}

The DASH library\footnote{\href{https://cosmo.oca.eu/dash/}{https://cosmo.oca.eu/dash/}} \citep{Ogiya2019} is a suite of idealized, collisionless $N$-body simulations that follow the evolution of an individual $N$-body subhalo as it orbits within the fixed, analytical potential of its host halo. Both the fixed host halo and the initial subhalo are spherically symmetric, each with a Navarro-Frenk-White \citep[NFW;][]{Navarro.etal.97} density profile:
\begin{equation}\label{eqn:nfw}
    \rho\sub{NFW}(r) = \rho_0 \, \left(\frac{r}{r_\rms}\right)^{-1} \,
  \left(1 + \frac{r}{r_\rms}\right)^{-2}\,,  
\end{equation}
where the model parameters $r_\rms$ and $\rho_0$ are the characteristic scale radius and density, respectively. The halo virial radius $\rvir$ is defined to be the radius within which the average density is $\Delta\sub{vir}=200$ times the critical density of the Universe $\rho\sub{crit}$. The corresponding virial mass is defined as $M\sub{vir} = \frac{4\pi}{3}\Delta\sub{vir}\rho\sub{crit} \rvir^3$. The halo concentration is defined as $c \equiv \rvir / r\sub{s}$, and the virial velocity as $V\sub{vir} \equiv \sqrt{GM\sub{vir}/\rvir}$. Throughout this work, the subscripts `h' and `s' represent quantities associated with the host- and subhaloes, respectively.

The initial conditions are generated assuming that the NFW subhalo has an isotropic velocity distribution, such that the phase-space distribution function (DF) depends only on energy. The simulations are performed with a tree code \citep{Barnes.Hut.86} developed for graphics processing unit (GPU) clusters \citep{Ogiya2013}. Each subhalo is initially made up of {1,048,576} particles, forces are softened with a Plummer equivalent length $\epsilon=0.0003r\sub{vir,s}$, and the opening angle of the tree is set to $\theta=0.7$. Orbits are integrated with the second-order leapfrog scheme with a global, adaptive time step $\Delta t = \sqrt{\epsilon/a\sub{max}}$, with $a\sub{max}$ the maximum, absolute acceleration among all particles at that time. As demonstrated in \citet{vdBosch.etal.2018b}, these parameters are sufficient to properly resolve the subhalo evolution.

For each simulation, the library contains various data about the subhalo evolution at 301 snapshots, with a physical time interval between each of 0.12\,Gyr. This corresponds to a total evolution time of 36\,Gyr, or 2.5 to 12 radial periods depending on the orbital configuration. The subhalo is initially placed at the apocenter of its orbit. At each timestep, DASH contains the radial profiles of the subhalo density, enclosed mass, and radial/tangential velocity dispersion, as well as its bulk position, velocity, bound mass fraction $f\sub{b}(t)$, and half-mass radius $r_h (t)$ (see Appendix A of \citet{vdBosch.etal.2018a} for details on how these quantities are computed). The radial profiles are computed for 40 logarithmically-spaced radial bins, which span $-2.95 \leq \log(r/r\sub{vir,s})\leq 0.95$.  While all DASH simulations initially meet the numerical reliability criteria of \citet{vdBosch.etal.2018b}, the simulations can become unreliable as the bound mass fraction becomes small. In this work, we only consider simulation snapshots that meet the following two reliability criteria, introduced in \citet{vdBosch.etal.2018b}, each of which can be computed using $r_\rmh (t)$ and $f\sub{b}(t)$. The first criterion, motivated by \citet{Power.etal.03}, demands that the softening length be sufficiently small to resolve the maximum particle accelerations, a requirement given by
\begin{equation}\label{eqn:crit1}
f_\rmb (t) > 1.79 \, \frac{\csub^2}{f(\csub)} \, \left(\frac{\epsilon}{\rsub}\right) \, \left(\frac{r_\rmh(t)}{\rsub}\right) .
\end{equation}
The second criterion, related to discreteness noise, states that the number of bound particles in the subhalo must exceed ${N\sub{crit}=80N^{0.2}}$, with $N$ the initial number of particles in the subhalo. Once the bound particle count falls below this value, the subhalo experiences a discreteness-driven runaway instability resulting in artificial disruption. In the DASH database, this requirement translates to
\begin{equation}\label{eqn:crit2}
    f\sub{b}(t) = 1.22\times 10^{-3} .
\end{equation}
We note that over $99.5\%$ of the DASH simulation snapshots meet the requirements of equations (\ref{eqn:crit1}) and (\ref{eqn:crit2}).

In addition to excluding snapshots that do not meet the numerical reliability criteria, we also perform several additional preprocessing steps. We exclude snapshots that are within the 10\% of the orbital period centered around pericentric passage in order to avoid intervals where $f_\rmb(t)$ is changing rapidly and the boundedness designation of individual particles is less reliable.\rev{\footnote{When this selection criterion is removed, our results remain qualitatively the same and we find that the variance in the residuals between our best-fit model and the DASH density profiles (as in Fig. \ref{fig:res_plot}) increases slightly at large subhalo radii.}} Additionally, only subhalo radial density profile points in the range $0.01\leq r/r\sub{vir,s} \leq 1$ are used for analysis; this innermost radius corresponds to ${\sim}3$ times the softening length, inside of which the density profile is not reliable.

The database contains 2,253 simulations of subhaloes orbiting within host haloes with an initial host-to-subhalo mass ratio of $M\sub{vir,h} / M\sub{vir,s} = 1000$, a ratio sufficiently large that the effects of dynamical friction \citep{Chandrasekhar.43} can safely be neglected. Furthermore, due to the self-similar nature of subhalo evolution, the simulations apply generally to initial configurations with $M\sub{vir,h} / M\sub{vir,s} \gtrsim 100$, regardless of the absolute value of $M\sub{vir,h}$. The simulations spread a four-dimensional parameter space of host- and subhalo concentrations and initial orbital configurations, as illustrated by Figs. 2 and 4 in \citet{Ogiya2019}. The concentrations $c\sub{h}$ and $c\sub{s}$ cover the range $3.1 \leq c \leq 31.5$, with the majority of the simulations devoted to the host- and subhalo concentrations (and ratios between the two) most commonly seen in cosmological simulations for haloes roughly in the range of $10^7 < M\sub{vir}/(h^{-1}M_\odot) < 10^{15}$, determined using the method described in Section 2.2.3 of \citet{Ogiya2019}. The initial orbital configuration is parametrized by two dimensionless analogs to energy and angular momentum: $x_\rmc \equiv r_\rmc(E)/r\sub{vir,h}$, where $r_\rmc(E)$ is the radius of the circular orbit of energy $E$, and the circularity $\eta=L/L_\rmc (E)$, where $L$ is the initial orbital angular momentum and $L_\rmc (E)$ is the angular momentum of the corresponding circular orbit with the same energy. The orbital parameters are sampled in the range $0 \leq \eta \leq 1$ (linearly) and $0.5 \leq x_\rmc \leq 2$ (logarithmically). The majority of the simulations are devoted to orbital parameters near the peak of the probability distribution seen at infall in cosmological simulations \citep{Jiang.etal.15}.

\section{Evolved Subhalo Density Profile}\label{sec:esdp}

The objective of this paper is to calibrate a model of the evolution of the subhalo density profile against the DASH simulations. As described above, the DASH database consists of 2,253 simulations, each of which has 301 snapshots of time evolution over several orbital periods.
At each of these snapshots, various radial profiles and global subhalo properties are stored. After performing the preprocessing steps described previously, the calibration dataset consists of a total of roughly $6\times 10^5$ snapshots of subhalo evolution labeled by (i) the initial configurations, which span the parameter space of $c\sub{h}$, $c\sub{s}$, $x\sub{c}$, and $\eta$ values, and by (ii) the bound fractions $f\sub{b}(t)$, which span roughly three orders of magnitude (${\sim}10^{-3}$ to $1$).  At each of these snapshots, we compute the ratio of the evolved subhalo density profile relative to the initial subhalo density profile, which we refer to as the {\it transfer function} $H(r,t) = \rho(r,t) / \rho(r,t=0)$, where $\rho(r,t=0)$ is the NFW profile of equation~(\ref{eqn:nfw}). The transfer function is stored for 20 radial bins spanning $0.01\leq r/r\sub{vir,s} \leq 1$ at each snapshot. This calibration dataset is immense, including over 10 million distinct data points of subhalo transfer functions.

The studies of \rev{\citet[][hereafter \citetalias{Hayashi.etal.03}]{Hayashi.etal.03}} and \citet{Penarrubia2008} argued that the subhalo density profiles depend solely on the density profile at infall and the total amount of mass lost thereafter. In particular, \citetalias{Hayashi.etal.03} describes the evolved density profile in terms of a transfer function, $H_\mathrm{H03}(r|f_\mathrm{b})$, which implies that the density profiles of subhaloes are insensitive to how and when they have lost their mass. Based on the same principle, \rev{\citet[][hereafter \citetalias{Penarrubia.etal.10}]{Penarrubia.etal.10}} provides a prescription to obtain a transfer function based off of their ``tidal track'' fitting function for the structural parameters normalized by their initial values, $\frac{V_\mathrm{max}}{V_\mathrm{max,i}}(f_\mathrm{b})$ and $\frac{r_\mathrm{max}}{r_\mathrm{max,i}}(f_\mathrm{b})$. Here $V_\mathrm{max}$ is the maximum circular velocity and $r_\mathrm{max}$ is the associated radius. Based on the DASH database, though, we find that the residuals between these models and the DASH transfer functions exhibit a significant, systematic correlation with the initial subhalo concentration, $c_\mathrm{s}$. Neither \citetalias{Hayashi.etal.03} nor \citetalias{Penarrubia.etal.10} observed this dependence, as both works only considered subhaloes with a single value for the concentration ($c_\rms=10$ and $23.1$, respectively).
\rev{In addition, we find that the dependence on $c_\rms$ is much stronger than on any of $c_\rmh$, $x_\rmc$, or $\eta$, which illustrates that while the evolved subhalo density profile depends on both the total amount of mass lost since infall \textit{and} the initial profile (encoded by $c_\rms$), the evolution is indeed independent of the details of the stripping (which depends on the external potential, encoded by $c_\rmh$, and the subhalo's orbit, encoded by $x_\rmc$ and $\eta$).}

Both \citetalias{Hayashi.etal.03} and \citetalias{Penarrubia.etal.10} find that tidal evolution modifies the subhalo density profile in two main ways:  (i) the outer density profile begins to drop off much more steeply with radius, transitioning from the $\rmd\log\rho/\rmd\log r = -3$ that is characteristic of the NFW profile at infall to $\rmd\log\rho/\rmd\log r = -(5-6)$, and (ii) the central densities slowly decrease with time as more and more mass is stripped away. The latter is mainly a consequence of the subhalo re-virializing in response to its mass loss. In addition, some of the reduction in central density arises more directly from the stripping of particles on highly eccentric orbits, which contribute mass to both the center and the outskirts. The impact of tidal shocking on the central densities is negligible as the short dynamical times in the dense centers imply adiabatic shielding \citep{Gnedin.Ostriker.99, vdBosch.etal.2018a}. Informed by these previous findings, and considering the newly-identified $c_\mathrm{s}$-dependence, we seek to describe the evolution of the subhalo density profile in terms of a transfer function $H(r | f_\mathrm{b}, c_\mathrm{s})$ that depends both on the initial subhalo concentration and the fraction of mass that has been stripped since infall.

Thus, the model-building procedure is largely one of exploratory data analysis and optimization. For calibrating candidate models of $H(r|f_\mathrm{b},c_\mathrm{s})$, we employ a cost function that is the sum of squared logarithmic residuals between the DASH transfer functions and those predicted by the model:
\begin{equation}\label{eqn:cost}
\begin{split}
        E(\bmath{\theta}) = \sum_{i}^{N_{\rm sim}} \sum_{j}^{N_{\rm snap}} \sum_{k}^{N_{\rm rad}} \big\{&\log\big[H_\rmD(r_k | t_j, \{c_\rmh, c_\rms, x_\rmc, \eta\}_i)\big] \\
        - &\log \big[ H_\rmm(r_k | f_\rmb(t_j), c_\rms, \bmath{\theta})\big]\big\}^2
\end{split}
\end{equation}
Here, $\bmath{\theta}$ denotes the free parameters of the model, and the sums run over all $N_{\rm sim}$ simulations, $N_{\rm snap}$ snapshots, and $N_{\rm rad}$ radial bins included in the preprocessed calibration dataset. $H_\rmD$ denotes the DASH transfer functions, which are labeled by the orbital parameters and halo concentrations at infall, snapshot number, and radial bin. $H_\rmm$ denotes the model transfer function, which only depends on the radial bin, bound fraction, initial subhalo concentration, and free model parameters. The adaptive Nelder-Mead downhill simplex method \citep{Gao2012} is used for model optimization due to its reliability and generalization to high-dimensional parameter spaces.

The DASH database does not contain a flat distribution of simulations across $c_\rmh$, $c_\rms$, $x_\rmc$, and $\eta$, but rather consists of proportionally more simulations in the regions of parameter space that are more probable. Furthermore, the snapshots present in our calibration dataset do not contain a flat distribution in $f_\rmb$, as there are far fewer snapshots of subhaloes with low $f_\rmb$ than for the highest values. Thus, by using our flat cost function, which weights all radial bins and all snapshots equally, the calibrated model will perform best in the regions of parameter space that are most commonly found in cosmological simulations.

After testing a variety of functional forms for $H(r| f_\rmb, c_\rms, \bmath{\theta})$, we find that the transfer function is quite well described by
\begin{equation}\label{eqn:tf}
    H(r | f_\rmb, c_\rms, \bmath{\theta}) = \frac{\rho(r,t)}{\rho(r,t=0)} = \frac{f_{\rm te}}{1 + \big(\tilde{r}\big[\frac{\tilde{r}\sub{vir,s} - \tilde{r}_{\rm te}}{\tilde{r}\sub{vir,s}\tilde{r}_{\rm te}}\big]\big)^\delta} ,
\end{equation}
which is a generalized form of the transfer function used in \citetalias{Hayashi.etal.03}, which is given by $H_\mathrm{H03}(r|f_\rmb) = f_\mathrm{te}[1+(\tilde{r}/\tilde{r_\rms})^3]^{-1}$. Here, $\tilde{r} = r / r_\rms$, such that all radii that appear in the transfer function are normalized to the initial NFW scale radius. The transfer function model contains three parameters:
\begin{equation}\label{eqn:fte}
    f_{\rm te} = f_\rmb^{a_1 \big(\frac{c_\rms}{10}\big)^{a_2}} c_\rms^{a_3 (1-f_\rmb)^{a_4}},
\end{equation}
\begin{equation}
    \tilde{r}_{\rm te} = \tilde{r}_{\rm vir,s} f_\rmb^{b_1 \big(\frac{c_\rms}{10}\big)^{b_2}} c_\rms^{b_3 (1-f_\rmb)^{b_4}} \exp\Big[b_5 \big(\frac{c_\rms}{10}\big)^{b_6} (1-f_\rmb)\Big],
\end{equation}
and
\begin{equation}\label{eqn:delta}
    \delta = c_0 f_\rmb^{c_1 \big(\frac{c_\rms}{10}\big)^{c_2}} c_\rms^{c_3 (1-f_\rmb)^{c_4}}.
\end{equation}
\rev{These parametrizations were motivated based on power series expansions in $\log(f_\rmb)$ and $\log(c_\rms)$ for the logarithms of $f_{\rm te}$, $\tilde{r}_{\rm te}$, and $\delta$. Additional coupling between $f_\rmb$ and $c_\rms$ was added and the functional forms were further adjusted through trial and error in order to maximally reduce the cost function in equation \eqref{eqn:cost}.}

Clearly, $f_\mathrm{te}$ describes how the normalization of the inner density profile evolves.
The other two parameters describe the steepening of the outer density profile. The tidal truncation radius $\tilde{r}_\mathrm{te}$ is related to the radius where the power-law begins to transition from NFW to a steeper, tidally stripped profile.
The power-law slope at large radii is governed by $\delta$, such that
\begin{equation}
    H(r) \propto r^{-\delta} \implies \rho(r) \propto r^{-(3+\delta)} \quad \mathrm{for} \quad \tilde{r} \gg \tilde{r}_\mathrm{te}.
\end{equation}
This transfer function has several desirable, physically-motivated properties. Firstly, when $f_\rmb=1$, the transfer function is unity for all radii, which is consistent with the fact that no tidal evolution has occurred yet.  Furthermore, the truncation radius $r_\mathrm{te}$ starts at the virial radius and shrinks inwards only as the subhalo is tidally stripped.

\begin{figure}
    \centering
    \includegraphics[width=\linewidth]{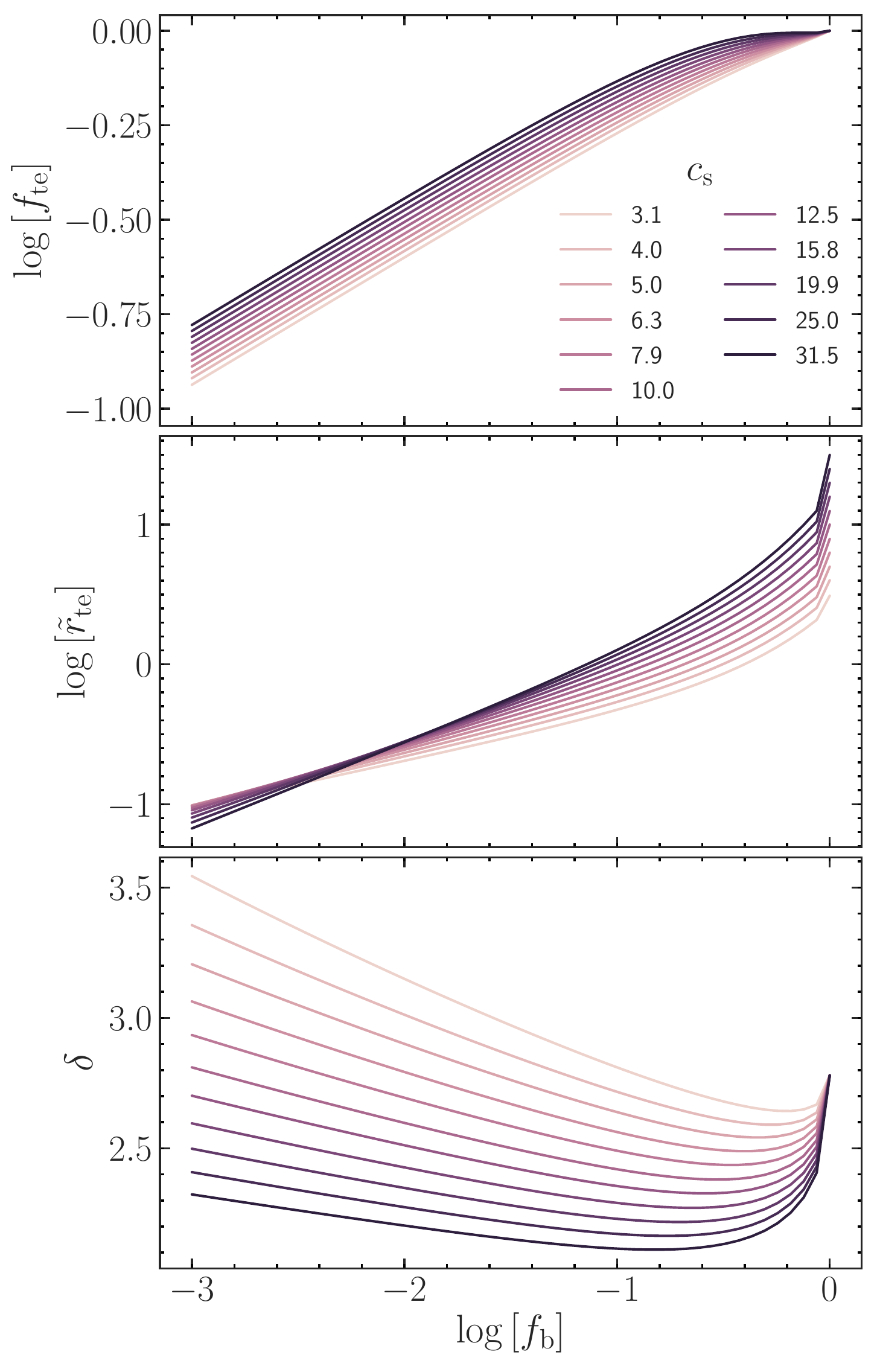}
    \caption{The dependence of the transfer function model (equation~[\ref{eqn:tf}]) functional parameters ($f_\mathrm{te}$, $\tilde{r}_\mathrm{te}$, and $\delta$, described by equations~[\ref{eqn:fte}]--[\ref{eqn:delta}]) on the subhalo concentration $c_\rms$ and bound fraction $f_\rmb$. For the majority of the $f_\rmb$-$c_\rms$ parameter space, $\delta\approx2-3$, resulting in a stripped subhalo density profile with $\rmd\log\rho/\rmd\log r = -(5-6)$, in agreement with previous idealized simulations \citepalias{Hayashi.etal.03,Penarrubia.etal.10}. \rev{As the subhalo is increasingly stripped, $\delta$ increases and the outer profile drops off more steeply.} Since $\rho(r) \propto f_\mathrm{te}$, the overall normalization of the density profile decreases as mass is stripped. The tidal truncation radius, $r_\mathrm{te}$, roughly corresponds to the radius where the profile transitions to $\rmd\log\rho/\rmd\log r = -(3+\delta)$; this radius is smaller for subhaloes that are initially more concentrated.}
    \label{fig:model_param_forms}
\end{figure}

Each of these three model parameters is itself parametrized to be a function of $c_\rms$ and $f_\rmb$. In total, the 15 free parameters to calibrate are encoded in $\bmath{\theta}$ as
\begin{equation}
    \bmath{\theta} = \{a_1, a_2, a_3, a_4, b_1, b_2, b_3, b_4, b_5, b_6, c_0, c_1, c_2, c_3, c_4 \} .
\end{equation}
We calibrate this model against the DASH simulations using the cost function and method described above, and the best fit parameters are listed in Table \ref{tab:free_params}. Additionally, the dependence of the three functional parameters, $f_\mathrm{te}$, $r_\mathrm{te}$, and $\delta$, on $f_\rmb$ and $c_\rms$ can be seen in Fig. \ref{fig:model_param_forms}. Importantly, unlike the polynomial expansions used in \citetalias{Hayashi.etal.03}, our power-law parametrizations of $f_\mathrm{te}$, $r_\mathrm{te}$, and $\delta$ are well-behaved down to arbitrarily low $f\sub{b}$. Such a property will be crucial for using the model in a semi-analytic prescription for evolving subhalo populations, which, in the absence of an explicit mechanism for subhalo disruption, will continue to evolve subhaloes down to $f\sub{b}$ below the resolution limit of DASH. \rev{For applications that do not depend on physically realistic extrapolation outside of the DASH $f\sub{b}$ parameter space, an alternative, promising strategy for predicting the evolved subhalo density profile could involve employing a machine learning algorithm, such as random forest regression \citep{Breiman2001}.} In agreement with previous works, our calibrated model demonstrates that the majority of the evolved subhalo density profiles are indeed well-described by $\rho \propto r^{-(5-6)}$ (i.e., $\delta \approx 2-3$). In particular, the outer density profile falls off more rapidly as subhalo concentration decreases.
\begin{table}
\begin{center}
\begin{tabular}{lS[table-format=-1.3]|lS[table-format=-1.3]|lS[table-format=-1.3]}
\hline
$a_1$  &  0.338  &  $b_1$  &  0.448  &  $c_0$  &  2.779  \\
$a_2$  &  0.000  &  $b_2$  &  0.272  &  $c_1$  & -0.035  \\
$a_3$  &  0.157  &  $b_3$  & -0.199  &  $c_2$  & -0.337  \\
$a_4$  &  1.337  &  $b_4$  &  0.011  &  $c_3$  & -0.099  \\
       &         &  $b_5$  & -1.119  &  $c_4$  &  0.415  \\
       &         &  $b_6$  &  0.093  &         &         \\
\hline
\end{tabular}
\caption{The best-fit parameters for the transfer function $H(r|f_\rmb, c_\rms)$ (equation~[\ref{eqn:tf}]). These parameters are used to describe the dependence of the model's functional parameters (i.e., $f_\mathrm{te}$, $\tilde{r}_\mathrm{te}$, and $\delta$, described by equations [\ref{eqn:fte}]--[\ref{eqn:delta}]) on the subhalo concentration $c_\rms$ and bound fraction $f_\rmb$. The best-fit value for parameter $a_2$ is consistent with zero, but the parameter was kept in order to maintain a consistent parametric form between $f_\mathrm{te}$ and the other functional parameters.}
\label{tab:free_params}
\end{center}
\end{table}

\begin{figure*}
    \centering
    \includegraphics[width=\linewidth]{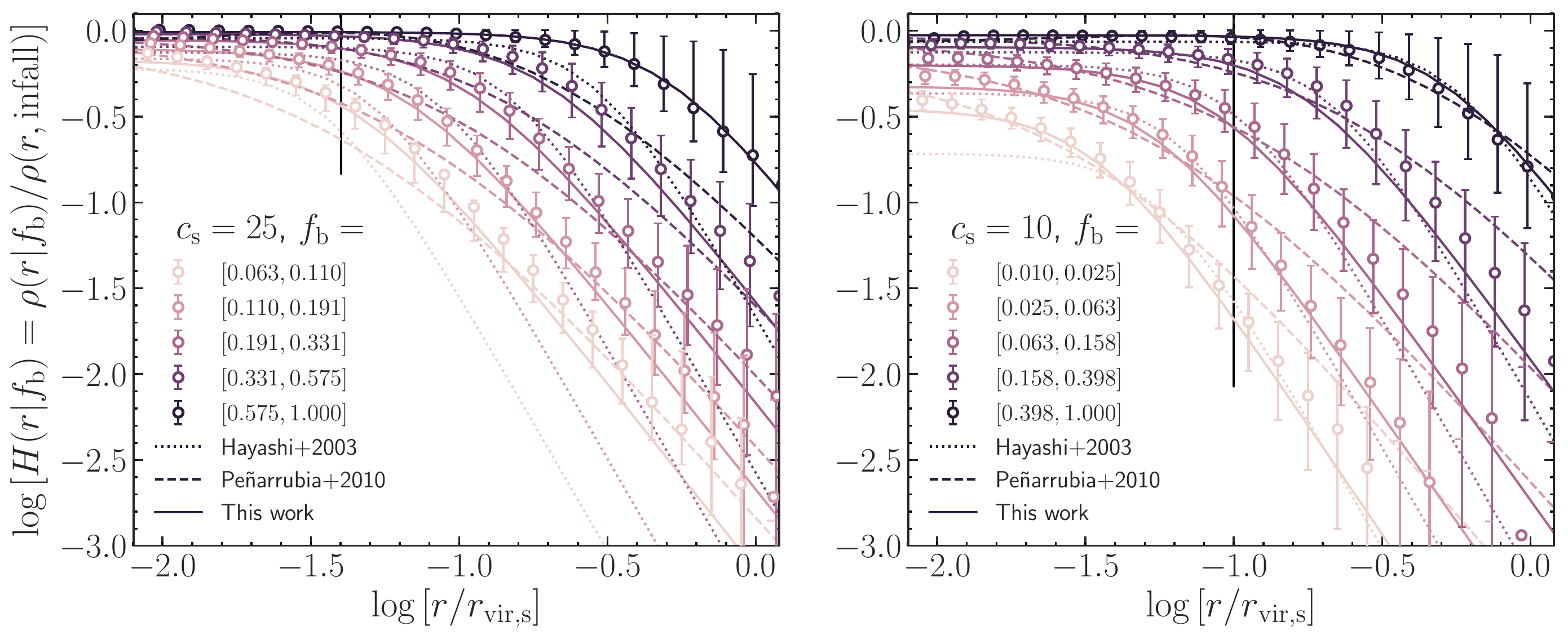}
    \caption{Comparison between evolved density profile models and the DASH simulations as a function of radius for $c_\rms=25$ ({\it left}) and $c_\rms=5$ ({\it right}), binned by $f_\rmb$. All DASH simulation snapshots for subhaloes with the specified $c_\rms$ that meet the preprocessing criteria (i.e., are numerically reliable and away from pericenter) and have $f_\rmb$ within the listed range are included. The open circles represent the median density profile transfer function value for the DASH simulations within the radial bin and $f_\rmb$ range, and error bars represent 16/84 percentiles. The $f_\rmb$ bins are progressively shifted horizontally for viewing in order to avoid overlapping error bars, but the true radii correspond to those of the lightest bin in $f_\rmb$. The solid vertical lines denote $r_\rms$. For a large range of $c_\rms$ values, the model accurately reproduces the tidally stripped subhalo density profile. The variable outer profile slope $\delta$, described by equation~(\ref{eqn:delta}), enables our model to better capture the outer density profile than \citetalias{Hayashi.etal.03} and \citetalias{Penarrubia.etal.10}, which use a fixed outer profile scaling of $\rmd\log\rho/\rmd\log r = -6$ or $\rmd\log\rho/\rmd\log r = -5$, respectively.}
    \label{fig:stacked_prof_comparison}
\end{figure*}

In Fig. \ref{fig:stacked_prof_comparison}, we compare our calibrated model to the DASH simulation transfer functions. Specifically, we first select a particular $c_\rms$, then bin the DASH simulation snapshots by $f_\rmb$, which includes simulations over the parameter space of $c_\rmh$, $x_\rmc$, and $\eta$ values. We plot these binned transfer functions versus radius, showing the medians and 16/84 percentiles for different ranges in $f_\rmb$, as indicated. Our model transfer function is specified by $c_\rms$, $f_\rmb$ (which is equal to the $f_\rmb$ logarithmic bin center used for the DASH data), and the radius. The model demonstrates good agreement with the DASH simulation transfer functions across a large dynamic range in $f_\rmb$ and over the relevant $c_\rms$ parameter space.  To highlight our improved model and emphasize the benefits of using a large library such as DASH for data-driven model building, we overplot the transfer functions of \citetalias{Hayashi.etal.03} and \citetalias{Penarrubia.etal.10}. As described above, these models for the transfer function depend only on $f_\rmb$. The model of \citetalias{Penarrubia.etal.10}, which was only calibrated to reproduce the structural parameters of subhaloes with $c_\rms=23.1$, is able to capture the outer density profile of highly-stripped subhaloes with $c_\rms=25$ quite well, whereas it fails to reproduce the corresponding inner density profiles. For the subhaloes with $c_\rms=10$, the \citetalias{Penarrubia.etal.10} model is better able to capture the inner density profile.  The model of \citetalias{Hayashi.etal.03}, which was calibrated only for subhaloes with $c_\rms=10$, performs better for low $c_\rms$, but is not able to capture the inner profile normalization as well as our model, especially for highly-stripped haloes. An accurate model of the subhalo transfer function needs to depend on the initial density profile (encoded by $c_\rms$), as is clear from the fact that both the models of \citetalias{Hayashi.etal.03} and \citetalias{Penarrubia.etal.10} perform much worse in the $c_\rms=25$ case than in the $c_\rms=10$ case. By incorporating dependence on $c_\rms$ into our transfer function model, we are able to better reproduce the DASH simulation transfer functions for both example initial subhalo concentrations. We also emphasize the benefit of using a variable outer power law ($\delta\approx 2-3$) for the transfer function. In most cases, the outer slope of our transfer function model is bracketed by the values advocated in \citetalias{Hayashi.etal.03} ($\delta=3$) and \citetalias{Penarrubia.etal.10} ($\delta=2$), enabling a more faithful reproduction of the outer profile across a broad range of $f_\rmb$ and $c_\rms$.

\begin{figure*}
    \centering
    \includegraphics[width=\linewidth]{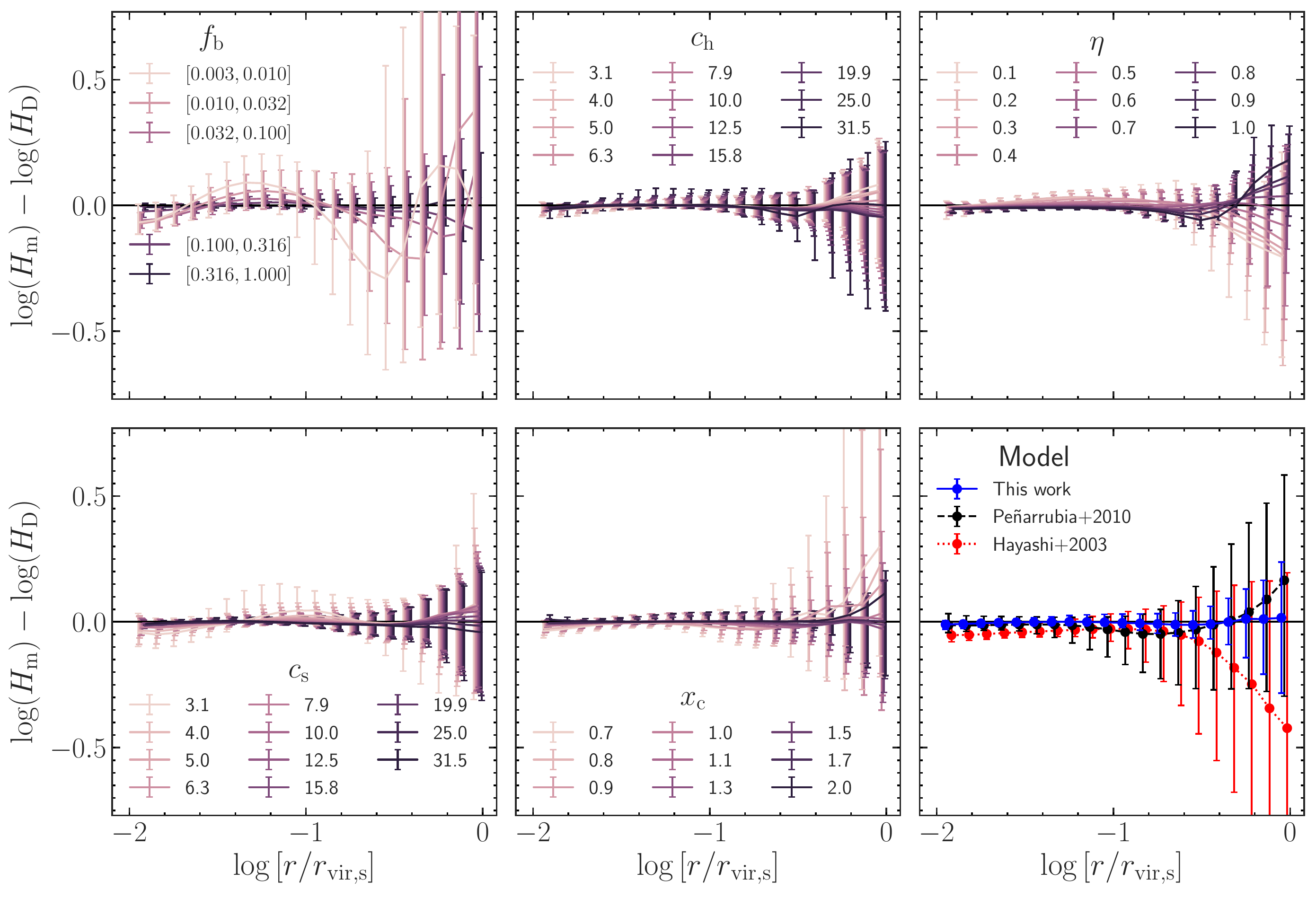}
    \caption{Residuals between the model and DASH simulation transfer functions ($H_\rmm$ and $H_\rmD$, respectively) as a function of radius, binned by $f_\rmb$ ({\it top left}), $c_\rms$ ({\it bottom left}), $c_\rmh$ ({\it top middle}), $x_\rmc$ ({\it bottom middle}), and $\eta$ ({\it top right}). The {\it bottom right} plot compares the residuals between our model and the DASH simulations (blue dots) to the residuals between each of the models of \citetalias{Hayashi.etal.03} (red dots) and \citetalias{Penarrubia.etal.10} (black dots) and the DASH simulations. Lines indicate the median residual and the error bars represent the 16/84 percentiles in each radial bin. The radii used for each value of the varied parameter (i.e., $f_\rmb$, $c_\rms$, $c_\rmh$, $x_\rmc$, $\eta$, or the model) are progressively shifted horizontally for viewing, but the true radii correspond to those of the lightest-coloured curves (or the curve corresponding to `this work' in the {\it bottom right} plot).
    See the text for a detailed discussion.}
    \label{fig:res_plot}
\end{figure*}

In Fig. \ref{fig:res_plot}, we plot the residuals between our model and the DASH simulation transfer functions, binned by radius and by each of $f_\rmb$, $c_\rms$, $c_\rmh$, $x_\rmc$, and $\eta$. We find that there is no significant systematic correlation between the residuals and $c_\mathrm{s}$ or $c_\mathrm{h}$. At the outer subhalo radii ($r\gtrsim 0.4 r_\mathrm{vir,s}$), the residuals increase for the most bound orbits (low $x_\rmc$) and exhibit a weak dependence on $\eta$. Note also that the model is least accurate for the lowest bound mass fractions (i.e., $f_\rmb \lta 0.01)$. Only a small fraction of all snapshots in DASH correspond to such small $f_\rmb$ values, all of which have small $c_\rms$. Consequently, this rare part of parameter space receives little weight in the optimization of the cost function, resulting in a less accurate fit. Note, though, that in each case the systematic offsets remain small compared to the halo-to-halo variance.

In the bottom right panel of Fig. \ref{fig:res_plot}, we give a final demonstration of the overall improvement of our model at reproducing the subhalo transfer functions of DASH compared to previous works. We plot the residuals between the various models and the DASH simulation transfer functions, now binned only by radius. These radial bins include all snapshots across the entire DASH dataset.  Clearly, our updated prescription for the transfer function significantly improves upon previous work, as demonstrated by its nearly negligible bias at all radii and substantially reduced scatter. In particular, the addition of a variable power-law slope in the transfer function eliminates the strong bias at large radii seen in the residuals of the other two models. Thus, our model, calibrated on a massive dataset that is less prone to the numerical artifacts that plague cosmological simulations, provides the best predictions to date for the evolution of the subhalo density profile. This tool will be a key ingredient in future semi-analytical models of dark matter substructure evolution.

\section{Structural Parameter Evolution}\label{sec:struct_evol}

Using the transfer function prescription developed above, one can easily compute the evolved subhalo density profile as $\rho(r | f_\rmb, c_\rms) = H(r | f_\rmb, c_\rms) \, \rho_\mathrm{NFW}(r | c_\rms)$. Using the evolved profile, the radius of the maximum circular velocity, $r_\mathrm{max}$, can be found by solving ${r^3 \rho(r) - \int_0^r r'^2 \rho(r')dr' = 0}$ for $r$. The associated maximum circular velocity is $V_\mathrm{max} = \sqrt{GM(<r_\mathrm{max}) / r_\mathrm{max}}$.

\citetalias{Penarrubia.etal.10} find that the structural parameters of subhaloes, $V_\mathrm{max}$ and $r_\mathrm{max}$, follow well-defined ``tidal tracks'' that only depend on $f_\rmb$ and the initial slope of the inner subhalo density profile. They calibrate a simple functional form for $V_\mathrm{max} / V_{\mathrm{max},i}(f_\rmb)$ and $r_\mathrm{max} / r_{\mathrm{max},i}(f_\rmb)$ based on their idealized subhalo simulations. They show that the functional form is accurate down to $f_\rmb \approx 0.001$ in their simulations. While their simulations span a variety of initial inner density profile slopes, all simulated subhaloes have $c_\rms=23.1$. On the other hand, \citetalias{Hayashi.etal.03}, who only analyzed idealized subhalo simulations with $c_\rms=10$, report that $V_\mathrm{max} \propto f_\rmb^{1/3}$; this result is inconsistent with the large cosmological simulations analyzed in \citet{Jiang.vdBosch.16} (see their Fig. 3) and, as we show below, is also inconsistent with DASH. The transfer function $H_\mathrm{H03}(r|c_\mathrm{s})$ of \citetalias{Hayashi.etal.03} can also be used to calculate the evolution of the subhalo structural parameters, yielding a different relation that is more consistent with other models and the DASH data.

In addition to reproducing the evolved subhalo density profile, the performance of the model can also be quantified by its ability to reproduce the evolved structural parameters. For the initial values, we use the structural parameters of an NFW halo: $V_{\mathrm{max},i}=0.465V_\mathrm{vir} \sqrt{c / f(c)}$ and $r_{\mathrm{max},i}=2.163r\sub{s}$ (here, $f(c) = \ln[1+c] - c/[1+c]$). In order to reduce the computational load of this analysis, we restrict ourselves to only the snapshots at apocentric passage, which still provides between 2--12 data points per simulation in the DASH database and a total of ${\sim}$9,000 snapshots. For each snapshot, we compute the empirical structural parameters using the enclosed mass profile stored in DASH. The circular velocity profile is computed for each radial bin as $V_\rmc (r) = \sqrt{GM(<r) / r}$ and then the structural parameters are determined from a fourth-order spline interpolation of this profile. Using each snapshot's associated values of $f_\rmb$ and $c_\rms$, the model predictions are calculated using the method described at the start of this section for our prescription and the one of \citetalias{Hayashi.etal.03}. The predictions of \citetalias{Penarrubia.etal.10} can be computed directly from their ``tidal track'' formula (their equation [8]).

\begin{figure*}
    \centering
    \includegraphics[width=\linewidth]{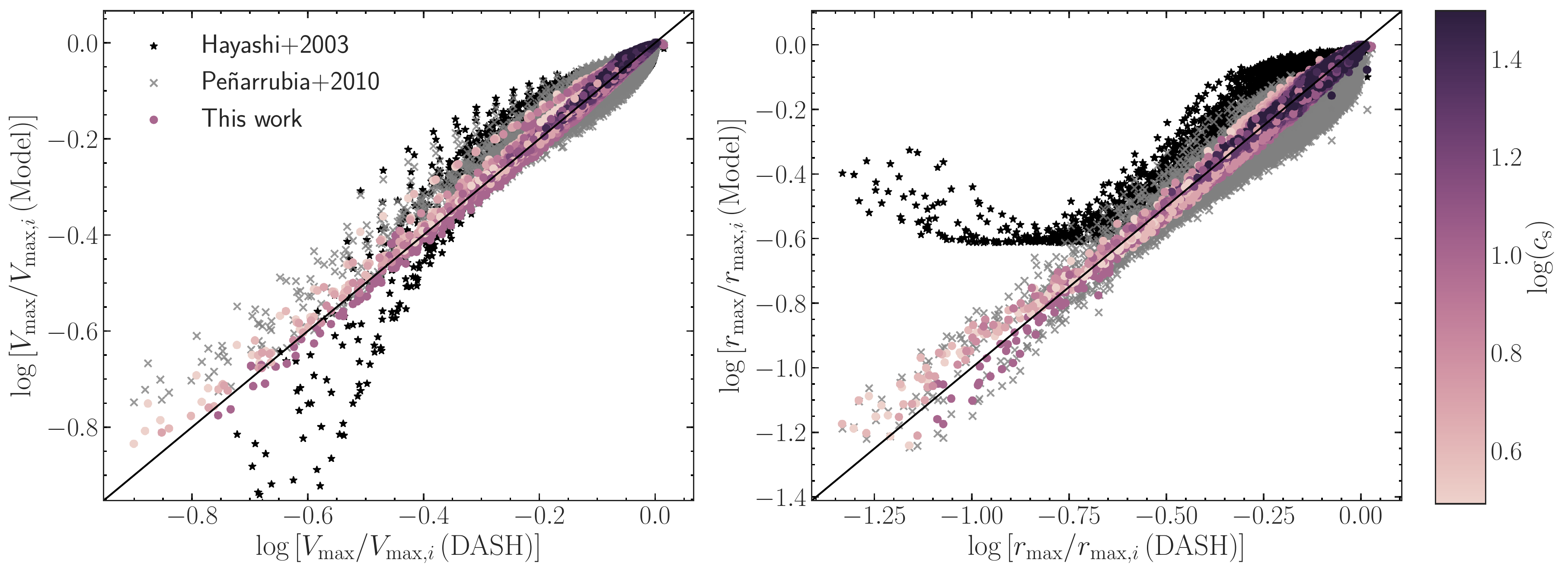}
    \caption{Scatter plots comparing the model predictions of the structural parameters normalized by their initial values, $V_\mathrm{max}/V_{\mathrm{max},i}$ ({\it left}) and $r_\mathrm{max}/r_{\mathrm{max},i}$ ({\it right}), to those of all DASH subhaloes at their apocentric passages. The results of our model are coloured by (the logarithm of) the subhalo concentration, demonstrating that the prediction's accuracy has minimal dependence on $c_\rms$. For comparison, the corresponding predictions from the models of \citetalias{Hayashi.etal.03} and \citetalias{Penarrubia.etal.10} are also plotted (black stars and gray crosses, respectively), highlighting their increased scatter.} 
    \label{fig:vmax_rmax_scatter}
\end{figure*}

\begin{figure*}
    \centering
    \includegraphics[width=\linewidth]{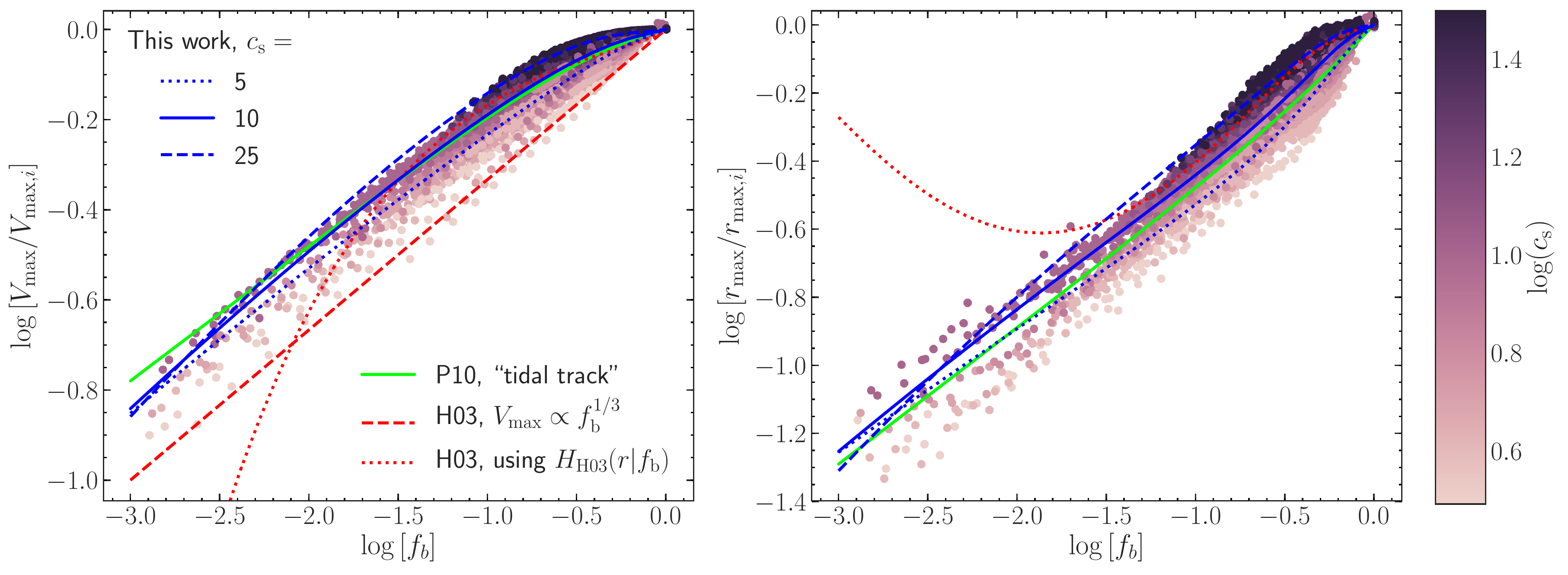}
    \caption{Scatter plots of $V_\mathrm{max}/V_{\mathrm{max},i}$ ({\it left}) and $r_\mathrm{max}/r_{\mathrm{max},i}$ ({\it right}) from all DASH subhalo snapshots at apocentric passages plotted against $f_\rmb$ and coloured by (the logarithm of) $c_\rms$. Overplotted are the model predictions of \citetalias{Hayashi.etal.03}, \citetalias{Penarrubia.etal.10}, and this work (equations~[\ref{eqn:mu}]-[\ref{eqn:eta}]). \citetalias{Hayashi.etal.03} reports that $V_\mathrm{max}\propto f_\rmb ^{1/3}$ and also provides a transfer function $H_\mathrm{H03}(r|c_\rms)$ that can be used to determine the structural parameters that results in a different relation. The latter are poorly behaved at small $f_\rmb$ due to the model's use of fitting functions that are polynomial expansions in $\log(f_\rmb)$. The \citetalias{Penarrubia.etal.10} predictions come directly from their ``tidal track'' fitting function (their equation [8]). The structural parameters can be determined using our transfer function model, which has dependence on $c_\rms$. As evidenced by the DASH data, such $c_\rms$-dependence is necessary in order to accurately capture the evolution of $V_\mathrm{max}$ and $r_\mathrm{max}$.}
    \label{fig:vmax_rmax_fb}
\end{figure*}

In Fig. \ref{fig:vmax_rmax_scatter}, we compare the model predictions for the structural parameters to the DASH results. Our model accuracy has minimal dependence on the stripped fraction, as evidenced by a similar level of scatter down to low $V_\mathrm{max}/V_{\mathrm{max},i}$ and $r_\mathrm{max}/r_{\mathrm{max},i}$. Additionally, the accuracy of our structural parameter predictions exhibits no residual dependence on the initial subhalo concentration. We overplot the predictions of \citetalias{Hayashi.etal.03} and \citetalias{Penarrubia.etal.10}, highlighting the significant improvement made by our model. In particular, much of the additional scatter in these prior models is due to the lack of $c_\rms$-dependence, which we illustrate below. In Fig. \ref{fig:vmax_rmax_fb}, we plot the DASH structural parameters against $f_\rmb$, coloured by the initial subhalo concentration. This plot demonstrates that at fixed $f_\rmb$, both $V_\mathrm{max}$ and $r_\mathrm{max}$ are larger for greater $c_\rms$, a trend that is exquisitely captured by our model due to the addition of $c_\rms$-dependence in the transfer function. A comparison between our model and the $H_\mathrm{H03}(r | f_\rmb)$-based structural parameter predictions illustrates the importance of using power law-based parametrizations in $f_\rmb$. By parametrizing the model's functional parameters ($f_\mathrm{te}$, $r_\mathrm{te}$, and $\delta$) as power laws in $f_\mathrm{b}$ and $c_\rms$, the transfer function and structural parameter predictions are well-behaved down to arbitrarily low $f_\rmb$, unlike the model of \citetalias{Hayashi.etal.03}, which uses a fitting function that is a polynomial expansion in $\log(f_\rmb)$.

Overall, our model's ability to accurately reproduce the evolved subhalo density profiles and associated structural parameters across a wide range of subhalo parameter space represents an important step towards building a more accurate model of dark matter substructure evolution.

In order to aid the building of such models, we provide additional fitting functions for $V_\mathrm{max}/V_{\mathrm{max},i}$ and $r_\mathrm{max}/r_{\mathrm{max},i}$. We use the same ``tidal track'' formula introduced in \citet{Penarrubia2008} and used in \citetalias{Penarrubia.etal.10}:
\begin{equation}\label{eqn:vm_rm_fit}
    X(f_\rmb,c_\rms) = \frac{2^\mu f_\rmb^\eta}{(1+f_\rmb)^\mu},
\end{equation}
where $\mu=\mu(f_\rmb,c_\rms)$, $\eta=\eta(f_\rmb,c_\rms)$, and $X$ denotes either $V_\mathrm{max}/V_{\mathrm{max},i}$ or $r_\mathrm{max}/r_{\mathrm{max},i}$. \citetalias{Penarrubia.etal.10} fit constants to each of the two functional parameters, $\mu$ and $\eta$; we introduce dependence on both $f_\rmb$ and $c_\rms$ and instead write them as:
\begin{equation}\label{eqn:mu}
    \mu(f_\rmb,c_\rms) = p_0 + p_1 c_\rms^{p_2} \log(f_\rmb) + p_3 c_\rms^{p_4},
\end{equation}
and
\begin{equation}\label{eqn:eta}
    \eta(f_\rmb,c_\rms) = q_0 + q_1 c_\rms^{q_2} \log(f_\rmb).
\end{equation}
The free parameters, $\bmath{p}$ and $\bmath{q}$, are fit to reproduce our model results for each of $V_\mathrm{max}/V_{\mathrm{max},i}$ and $r_\mathrm{max}/r_{\mathrm{max},i}$; the resulting values are listed in Table \ref{tab:vm_rm_free_params}. The fitting function agrees with our model to $\lesssim 1\%$ for $V_\mathrm{max}/V_{\mathrm{max},i}$ and $\lesssim 3\%$ for $r_\mathrm{max}/r_{\mathrm{max},i}$ over the range $-3 \leq \log(f_\rmb) \leq 0$ and $3.1 \leq c_\rms \leq 31.5$. Both the full transfer function model and the structural parameter fitting functions are well-behaved down to arbitrarily low $f_\mathrm{b}$, which is a crucial characteristic for use in a semi-analytical model without disruption.
\begin{table}
\begin{center}
\begin{tabular}{lS[table-format=-1.3]|lS[table-format=-1.3]|lS[table-format=-1.3]|lS[table-format=-1.3]}
\hline
\multicolumn4c{$V_\mathrm{max}/V_{\mathrm{max},i}$} & \multicolumn4c{$r_\mathrm{max}/r_{\mathrm{max},i}$} \\
\hline
$p_0$  &  2.980  &  $q_0$  &  0.176  &  $p_0$  &  1.021  &  $q_0$  & -0.525  \\
$p_1$  &  0.310  &  $q_1$  & -0.008  &  $p_1$  &  1.463  &  $q_1$  & -0.065  \\
$p_2$  & -0.223  &  $q_2$  &  0.452  &  $p_2$  &  0.099  &  $q_2$  &  0.083  \\
$p_3$  & -3.308  &         &         &  $p_3$  & -4.643  &         &         \\
$p_4$  & -0.079  &         &         &  $p_4$  & -0.250  &         &         \\
\hline
\end{tabular}
\caption{The parameters of the fitting function for the subhalo structural parameters normalized by their initial values, $V_\mathrm{max}/V_{\mathrm{max},i}$ and $r_\mathrm{max}/r_{\mathrm{max},i}$ (equation [\ref{eqn:vm_rm_fit}]), calibrated to agree with our transfer function model. These parameters encode the dependence of the model's functional parameters, $\mu$ and $\eta$, on the subhalo concentration $c_\rms$ and bound fraction $f_\rmb$ (equations~[\ref{eqn:mu}]--[\ref{eqn:eta}]).}
\label{tab:vm_rm_free_params}
\end{center}
\end{table}

\section{Summary and Discussion}\label{sec:discussion}

The evolution of dark matter haloes is predominantly studied through cosmological $N$-body simulations. These simulations show that haloes in virial equilibrium have universal density profiles \citep[e.g.,][]{Navarro.etal.97} and maintain a population of subhaloes that contain roughly 10\% of the total halo mass \citep[e.g.,][]{Ghigna.etal.98,Gao.etal.04,Giocoli.etal.10}. It has been shown that a large fraction of such subhaloes present in these simulations are completely disrupted within only a few orbital periods \citep{Han.etal.16,vdBosch.17}.  Recently, several works have employed a combination of physical arguments and idealized simulations to claim that much of this subhalo disruption is artificial \citep{Penarrubia.etal.10,vdBosch.etal.2018a,vdBosch.etal.2018b}, indicating that the classical `over-merging' problem \citep[e.g.,][]{Katz.White.93, Moore.etal.96a} may still plague modern cosmological simulations. Specifically, \citet{vdBosch.etal.2018b} showed that artificial disruption is primarily due to discreteness noise and inadequate force softening, a numerical issue that has been able to elude standard convergence tests. Hence, alternative approaches to studying the statistics of dark matter substructure are essential in order to cross-check the results of state-of-the-art simulations; only this will guarantee our ability to extract maximum information content that can be used for constraining the nature of dark matter and furthering the small-scale cosmology program.

As a promising alternative to $N$-body simulations, the semi-analytical modeling approach combines analytical halo merger trees, built using extended Press-Schechter theory \citep{Bond.etal.91}, with a prescription for the tidal evolution of individual subhaloes as they orbit their host. This approach has been employed in a variety of previous models of substructure evolution \citep{Taylor.Babul.01, vdBosch.etal.05, Penarrubia.Benson.05, Zentner.etal.05, Diemand.etal.07a, Kampakoglou.Benson.07, Gan2010, Pullen.etal.14, Jiang.vdBosch.16}. These benefit from not being directly obstructed by the same numerical issues present in cosmological simulations. However, due to the lack of a fully analytical description of tidal evolution, these models still must be calibrated in some way against cosmological simulations (hence \textit{semi}-analytical). The free parameters of the model are typically determined by tuning the results to reproduce the empirical subhalo mass functions of cosmological simulations. Clearly, if a large fraction of subhaloes in the simulations are subject to spurious disruption, then the semi-analytical models are calibrated against artificially suppressed subhalo mass functions, ultimately inheriting the same inadequacies of the simulations.

In an attempt to circumvent this issue, \citet{Ogiya2019} introduced the DASH subhalo evolution database, a suite of 2,253 idealized, high-resolution $N$-body simulations of individual subhaloes orbiting within a static, analytical host halo. These simulations are unimpaired by artificial disruption, with over 99.5\% of the roughly $6\times 10^5$ snapshots in the database passing the conservative numerical reliability criteria of \citet{vdBosch.etal.2018b}. The library samples the entire region of parameter space (i.e., initial orbital configurations and host-/subhalo concentrations) consistent with dark matter substructure observed in cosmological simulations.

This work represents the first phase of a research program devoted to building a semi-analytical model of dark matter substructure evolution that is calibrated against the DASH database and thus unobstructed by artificial disruption. In particular, this program will enable a calculation of the evolved subhalo mass function that is entirely independent of cosmological simulations, yielding a powerful method for validating the (small-scale) results of such simulations. In this paper, we present an updated prescription for the evolution of the subhalo density profile. Previous such models by \citetalias{Hayashi.etal.03} and \citetalias{Penarrubia.etal.10} only depend on the fraction of matter that has become unbound from the subhalo since infall (described by $f_\rmb$). We find that the residuals between these $f_\rmb$-only models and the DASH subhalo density profiles correlate significantly with the subhalo concentration $c_\rms$. Hence, we propose a more general model that depends both on $f_\rmb$ and the initial profile at infall (described by $c_\rms$). This evolved subhalo density profile is described by the transfer function $H(r | f_\rmb, c_\rms) = \rho(r,t) / \rho(r,t=0)$, where we assume $\rho(r,t=0) = \rho_\mathrm{NFW}(r)$. Our model of this transfer function can be easily implemented in future semi-analytical models, as it has a simple algebraic form and is described fully by a set of parameters calibrated against the DASH simulations (see equations~[\ref{eqn:tf}]-[\ref{eqn:delta}] and Table \ref{tab:free_params}). As demonstrated in \S\ref{sec:esdp} and \S\ref{sec:struct_evol}, our model is able to reproduce far more accurately the density profiles and structural parameters of evolved subhaloes than the models of previous work. In addition, we provide a fitting function for the evolving structural parameters, described by equations \eqref{eqn:vm_rm_fit}--\eqref{eqn:eta} and Table \ref{tab:vm_rm_free_params}.

In the next paper in this series (Green et al., in prep.), we utilize the DASH library and our prescription for the evolved subhalo density profile to build a simple, physically-motivated model of the mass evolution of dark matter subhaloes. We will then combine this subhalo evolution model with accurate halo merger trees \citep[e.g.,][]{Parkinson.etal.08,Jiang2014} to predict the evolved subhalo mass function of CDM haloes, a result that is completely free from the effects of artificial disruption. This will allow us to verify the predictions of Fig. \ref{fig:shmf} and determine whether or not the subhalo mass and velocity functions have indeed been severely underestimated.
The results of this upcoming work will serve as an important check on the reliability of subhalo statistics derived from state-of-the-art cosmological simulations.

\section*{Acknowledgements}

The authors thank Fangzhou Jiang for supplying the data used for Fig.~\ref{fig:shmf}. SBG is supported by the US National Science Foundation Graduate Research Fellowship under Grant No. DGE-1752134. FCvdB is supported by the National Aeronautics and Space Administration through Grant No. 17-ATP17-0028 issued as part of the Astrophysics Theory Program and by the Klaus Tschira foundation.

\bibliographystyle{mnras}
\bibliography{references_vdb}

\bsp	
\label{lastpage}
\end{document}